\documentclass[twocolumn, amsmath,amssymb, superscriptaddress]{revtex4}

\usepackage{dcolumn}
\usepackage{bm}
\usepackage{amsmath}
\usepackage{amsfonts}
\usepackage{mathrsfs}  
\usepackage{graphics}
\usepackage[dvips]{graphicx}
\usepackage{times}
\usepackage{setspace}
\usepackage{color}   
\usepackage{verbatim}
\usepackage[raggedright]{titlesec}
\usepackage[normalem]{ulem}
\usepackage{framed}
\usepackage{enumerate}
\usepackage{footmisc}
\usepackage[breaklinks=true, hyperfootnotes=false]{hyperref}
\usepackage{breakurl}

\usepackage[graphicx]{realboxes}
\usepackage{varwidth}
\usepackage{footmisc}
\usepackage{balance}
\usepackage{flushend}

\hypersetup{colorlinks=true,citecolor=blue, linkcolor=black,urlcolor=blue} 

\graphicspath{{Figures/}}
\usepackage{hyperref}
\usepackage{amssymb}
\usepackage{mathbbol}
\usepackage{flushend}
\usepackage[outercaption]{sidecap}

\begin{document}
\title{On the social and cognitive dimensions of wicked environmental problems \\ characterized by conceptual and solution uncertainty}
\author{Felber J. Arroyave$^{a}$}
\affiliation{Department of Management of Complex Systems, Ernest and Julio Gallo Management Program,\\ School of Engineering, University of California, Merced, California 95343, USA}

\author{Oscar Yandy Romero Goyeneche$^{a}$}
\affiliation{Utrecht University Centre for Global Challenges, the Netherlands}

\author{Meredith Gore}
\affiliation{Department of Geographical Sciences, University of Maryland, College Park, MD, 20742 USA}

\author{Gaston Heimeriks}
\affiliation{Copernicus Institute of Sustainable Development, Utrecht University, the Netherlands}

\author{Jeffrey Jenkins}
\affiliation{Department of Management of Complex Systems, Ernest and Julio Gallo Management Program,\\ School of Engineering, University of California, Merced, California 95343, USA}

\author{Alexander M. Petersen}
\affiliation{Department of Management of Complex Systems, Ernest and Julio Gallo Management Program,\\ School of Engineering, University of California, Merced, California 95343, USA}



\maketitle

\footnotetext[1]{\ \ $^{a}$ These authors contributed equally. Send correspondence to: farroyavebermudez@ucmerced.edu  or apetersen3@ucmerced.edu}

{\bf \noindent We develop a quantitative framework for understanding the class of wicked problems that emerge at the intersections of natural, social, and technological complex systems. Wicked problems reflect our incomplete understanding of interdependent global systems and the systemic risk they pose; such problems escape solutions because they are often ill-defined, and thus mis-identified and under-appreciated by communities of problem-solvers. While there are well-documented benefits to tackling boundary-crossing problems from various viewpoints, the integration of diverse approaches can nevertheless contribute confusion around the collective understanding of the core concepts and feasible solutions. We explore this paradox by analyzing the development of both scholarly (social) and topical (cognitive) communities -- two facets of knowledge production studies here  that contribute towards the evolution of knowledge in and around a problem, termed a knowledge trajectory -- associated with three wicked problems: deforestation, invasive species, and wildlife trade. We posit that saturation in the dynamics of social and cognitive diversity growth is an indicator of reduced uncertainty in the evolution of the comprehensive knowledge trajectory emerging around each wicked problem. Informed by comprehensive bibliometric data capturing both social and cognitive dimensions of each problem domain, we thereby develop a framework that assesses the stability of knowledge trajectory dynamics as an indicator of wickedness associated with conceptual and solution uncertainty. As such, our results identify wildlife trade as a wicked problem that may be difficult to address given recent instability in its knowledge trajectory. }\\

\vspace{-0.1in}
Scientific knowledge production is a necessary input for better responding to the direct consequences, downstream impacts, and systemic risk associated with complex problems \cite{bettencourt_evolution_2011, bloomfield_politics_2020, defries_ecosystem_2017, helbing2013globally, patterson_exploring_2017}. Environmental problems in particular, such as climate change or biodiversity loss, happen at the multidisciplinary intersection of ecological, social and technological systems, and are therefore inherently complex \cite{ramirez2020fostering, smith2008social}. Managing such challenging boundary-spanning problems calls on the convergence of knowledge and expertise across disciplines and inter-sectoral organizations \cite{petersen_grand_2021}. Although different studies have addressed how knowledge is produced and integrated within well-stablished disciplinary domains (e.g., astronomy), little is known regarding scientific knowledge production about emerging environmental problems within the broad envelope of sustainability science \cite [e.g.,][]{bettencourt_evolution_2011}.

In particular, we are motivated by \emph{wicked problems}, typified as untamed, dynamically complex, and ill-structured problems, that lack clear-cut conceptual and solution definition \cite{rittel_dilemmas_1973,alford_wicked_2017, camillus_strategy_2008,head_wicked_2008}. Note that not all problems suffer the same degree and type of ill-definition, or are equally `wicked', as we further discuss. These known but largely unattended problems \cite{wucker_gray_2016} are elusive given the multiple interdependencies and the absence of a 'correct' view \cite{batie_wicked_2008}.

In order to fully appreciate the nature and implications of wicked problems, communities of problem-solvers are needed in convergence to bridge knowledge across disciplines, thereby triggering a common vision regarding the properties of the core problems, and the means to address and manage them \cite{alford_wicked_2017}. Such knowledge integration depends upon blending existing concepts and social structures (e.g., formal communities of researchers) that collectively and intentionally delineate an encompassing \emph{knowledge trajectory} that defines the scope of knowledge around a given domain \cite{heimeriks_how_2016}. Knowledge trajectories are thus defined as the characteristic dynamical pathway followed by a domain in both its cognitive (i.e., ideas and concepts) and social dimensions \cite{boschma_scientific_2014, heimeriks_how_2016, wagner_network_2005, bonaccorsi2008search}. Although the study of knowledge trajectories has advanced towards identifying how knowledge converge, diverge and eventually stabilize \cite[e.g.,][]{bonaccorsi2008search, bettencourt_scientific_2009, bettencourt_evolution_2011, boschma_scientific_2014, heimeriks_how_2016}, few studies have explored how knowledge trajectories emerge within the scope of ill-defined or wicked problems, where paradoxically, the multiple approaches can nevertheless contribute confusion around the collective understanding of the core concepts and feasible solutions. 

Against this backdrop, we seek to provide a better understanding of knowledge production around wicked problems starting with the question: \emph{do  knowledge trajectories emerging around wicked environmental problems differ according to their cognitive and social dimensions?} As such, for a given problem domain, we seek to elaborate on the association between the emergence and stabilization of its knowledge trajectory, as it relates to the overall wickedness of the underlying problem. 

We develop this framework by constructing a representation of the knowledge trajectory for each of three environmental problems -- Deforestation, Invasive Species, and Wildlife trade. These three problem domains were selected due to the great risks for negative ecological and societal impact they pose, owing to manifest systemic-risk associated with interdependent natural, social and technological systems
\cite{defries_ecosystem_2017, helbing2013globally, stirling_keep_2010, challender2021mis, folke2007interdependent}. Specifically, a common criterion for selecting these three problems is the lack of technically well-posed objectives, as multiple disciplinary lenses might prioritize different components of the system and therefore the ways to address the problem. We argue that these three environmental problems lack certainty regarding the core concepts and possible solutions, and that such flaws are sufficient to give rise to instability in the knowledge trajectory dynamics which manifests in exacerbating conceptual and solution uncertainty.

This work contributes to the literature on the emergence and dynamics of collective knowledge production\cite{bettencourt_evolution_2011,bettencourt_scientific_2009,boschma_proximity_2005,dolfsma_lock-and_2009,heimeriks_how_2016,rafols_diversity_2010,sun_social_2013,petersen_multiscale_2018}. To assess the emergence and stability of the knowledge trajectory for each problem domain, we develop an empirical data-driven framework focusing on the diversity of topics, disciplines, collaboration, and geographic coordination. We associate each empirical facet with either (a) cognitive dimension or (b) the social dimension. By simultaneously comparing networks representing (a) and (b), in a similar vein to prior research characterizing knowledge domains \cite{shiffrin_mapping_2004, heimeriks_emerging_2012, petersen_grand_2021}, we seek to identify patterns of scientific knowledge production related to wicked problems. 

The remainder of this paper is structured as follows in Section \ref{sec:2} we describe the relationship between cognitive and social dimensions of a knowledge trajectory, in the particular context of wicked problems. Sections \ref{sec:3} introduces the data and methodological framework for assessing the proposed relationships. Section \ref{sec:4} presents our analysis on the cognitive and social dimensions of knowledge trajectories. Finally, we conclude by discussing how observed (in)stability in the social and cognitive dimensions relates to the 3-level typology of wicked problems developed by Heifetz \& Heifetz \cite{heifetz_leadership_1994}.

\vspace{-0.2in}
\section{Conceptual Background}
\vspace{-0.1in}
\label{sec:2}

\vspace{-0.1in}
\subsection*{Defining the characteristics of wicked problems} 
\vspace{-0.1in}
' Wicked' is a concept that emerged from public policy research (also frequently used in studies of science dynamics) referring to particular characteristics of a knowledge domain \cite{alford_wicked_2017, heifetz_leadership_1994}. However, we are unaware of literature exploring the relationship between `wickedness' and its implications on scientific knowledge production. Hence, we seek to develop statistical methods for measuring, evaluating and better understanding wicked problems.

Problems can be defined, among other notions, by the degree to which related clear-cut concepts and solutions are identifiable. Heifetz \& Heifetz \cite{heifetz_leadership_1994} propose that in respect to the baseline of tame problems (Type I), wicked problems can be divided between those with well-defined conceptual definitions but with ill-defined solutions (Type II); and those lacking both well-defined conceptual definitions and solutions. (Type III). 

In essence, not all wicked problems have the same degree of wickedness. On the one hand, Type II wicked problems are conceptually clear but appear `fuzzy' to problem solvers, as they lack a single exact solution, or alternatively, are faced with multiple  solution pathways characterized by  uncertainty \cite{casals_fuzzy_1986, defries_ecosystem_2017, head_wicked_2008, head2019forty}. On the other hand, Type III wicked problems are inherently resistant to clear and unique definitions \cite{alford_wicked_2017, camillus_strategy_2008, defries_ecosystem_2017}, and are characterized by definition and solution uncertainty  \cite[see,][]{batie_wicked_2008, head_wicked_2008, heifetz_leadership_1994, light_wicked_2020}. In contrast to tame problems, wicked problems result in thorny issues for which common top-down expert-driven approaches can be insufficient to cope with their complexity \cite{defries_ecosystem_2017, heifetz_leadership_1994, funtowicz_risk_1992,schot2018deep,rittel_dilemmas_1973}. In the present context, we acknowledge that environmental problems aren't simply differentiated as tame or wicked, but given their tendency to be situated at the nexus of interdependent complex systems, they are distributed in a spectrum of ill-definition or wickedness \cite{alford_wicked_2017,head_wicked_2008, head2019forty}.

Beyond the issues of ill-defined concepts and solutions, and the multiplicity of conceptual and solution approaches, wicked problems are also exacerbated by social factors. Indeed, problems derived from anthropogenic drivers are socially situated \cite{alford_wicked_2017, camillus_strategy_2008}. Hence, addressing wicked problems facing society and planet requires convergent research spanning traditional disciplinary boundaries that leverages cross-sectoral integration of expertise \cite{bammer_expertise_2020, council_convergence_2014, light_wicked_2020, linkov_scientific_2014, petersen_grand_2021,stirling_keep_2010}. Consequently, the variety of stakeholders, interests, and objectives engaged in the social context may involve a large collection of opinions and ideas about the problem itself and its causes that can hinder consensus formation around a shared vision \cite{alford_wicked_2017,batie_wicked_2008, camillus_strategy_2008, defries_ecosystem_2017, funtowicz_risk_1992, rittel_dilemmas_1973}. For this reason, it is commonly appreciated that the greater the disagreement among stakeholders, the more wicked the problem is likely to be. Confusion, discord, and lack of progress are telltale signs that an issue might be wicked \cite{camillus_strategy_2008}. However, different studies \cite [e.g.,][]{bettencourt_scientific_2009, petersen2015quanti, wagner_network_2005, wagner2011approaches} also indicate that long term social interactions between stakeholders is critical to the consequential diffusion of knowledge, second-order learning and co-production of stable agreements \cite{schot_three_2018, stirling_general_2007, schot2018deep}. And while some scholars argue that the multiplicity of stakeholders is the primary factor contributing to wicked problems, we argue that such multiplicity is not a sole determinant. Instead, we posit that social and cognitive integration plays a dominant role in fostering the consolidation of knowledge domains, policy agendas, and shared vision, which together can reduce conceptual and solution uncertainties 
\cite{bettencourt_evolution_2011, defries_ecosystem_2017, guo_mixed-indicators_2011, heimeriks_how_2016, rafols_diversity_2010,stirling_general_2007}. 

We seek to provide clarity around this point by analyzing cognitive and social dimensions of knowledge trajectories \cite{rafols_diversity_2010}. Regarding the first dimension, we posit that cognitive factors are most likely to obfuscate the clarity of problem definitions. Such lack of agreement around concepts and their relationships is characteristic of endeavors calling on multi-disciplinary problem-solving. Consequently, wicked problems may fail to consolidate into stable trajectories because new efforts fail to constructively leverage and contribute to existing knowledge \cite{boschma_scientific_2014, heimeriks_how_2016}. 

Regarding the second dimension, social factors tend to contribute uncertainty concerning the set of solution pathways. We posit that the benefits of collaboration are less potent in research communities lacking clearly delineated pathways forward, and conversely, that problems lacking identifiable pathways forward are less likely to elicit stable community formation. Indeed, establishing and sustaining consequential leadership  may be untenable in wicked scenarios if there is a failure to alert, activate, orient, and incentivize the vast field of candidate problem solvers \cite{rittel_dilemmas_1973}. Ideally, critical scientific agendas become institutionalized as ' Grand Challenges' that serve as a lighthouse beacon to guide trajectories toward a clearly identifiable objective \cite{helbing2013globally, petersen_grand_2021}. Another important consideration is that wicked problems are by definition intractable, thereby lacking a single 'closed-form' solution; hence, ' better' solutions are converged upon instead of a unique 'correct' one. Such a process is typically feasible when stakeholders iteratively converge in agreement on how to institutionalize agendas that best address the problem \cite{alford_wicked_2017, batie_wicked_2008, camillus_strategy_2008, funtowicz_risk_1992, funtowicz_uncertainty_1994, stirling_keep_2010}. As such, wicked problems are commonly managed, as opposed to being solved, giving rise to a situation that requires long-standing agreements between stakeholders, robust research agendas, and inter- and trans-disciplinary approaches -- e.g., as Masterson \cite{masterson_malaria_2014} shows for the case of managing the malaria crisis in global tropical and subtropical zones.

\vspace{-0.2in}
\subsection*{The relation between problems and knowledge trajectories} 
\vspace{-0.1in}
Various studies have addressed the knowledge trajectories for well-stablished domains (e.g., astrophysics; organic chemistry) identifying that they eventually reach stable trajectories reflecting the steady-state accumulation of intellectual development \cite[e.g.,][]{bonaccorsi2008search, heimeriks_how_2016, heimeriks_emerging_2012, wagner2011approaches}. Such stability emerges from the incremental addition of coherent knowledge; in other words, knowledge based upon the preexisting. Although emerging fields (e.g., computer sciences) show a more turbulent pattern characterized by fluctuations owing to disruptive innovations, and therefore less related knowledge. Such fields nevertheless eventually achieve stability in their richness, e.g. proxied by descriptors like the rate of new keywords used -- e.g., see Bonaccorsi \cite{bonaccorsi2008search}. Similarly, research on social dimensions \cite[e.g.,][]{bettencourt_population_2008, bettencourt_evolution_2011,petersen_multiscale_2018} show that social cohesion manifesting as consolidated collaboration is a common characteristic of synergistic cross-disciplinary integration; alternatively, if persistent collaboration is lacking, then it is less likely that consequential blending of concepts and methods will succeed. 

A common theme in knowledge trajectory research is what role relatedness plays in knowledge diversification \cite{boschma_scientific_2014, dolfsma_lock-and_2009, heimeriks_emerging_2012, heimeriks_how_2016}. Implied in this definition of relatedness is the strong path dependency regarding the entry and exit of knowledge building-blocks accumulated in the system \cite{heimeriks_how_2016}. We argue that cognitive relatedness thus captures to the permanence, continuity, and integration (with the preexisting) of concepts and ideas.

Similar to cognitive relatedness, social relatedness refers to the permanence and reinforcement of pre-existing associations between partners. Long-term collaborations underly established process of academic debate \cite{grauwin_mapping_2011, wagner_network_2005} that assists in consolidating agendas, enables deep learning within and across academic communities \cite{bettencourt_evolution_2011, granovetter_economic_1985, reinders_role_2011}, is associated with higher-impact research outcomes \cite{petersen2015quanti}. Stable collaboration is therefore an indicator of social consolidation of knowledge trajectories \cite{bettencourt_evolution_2011}.

Cognitive and social networks provide a well-established framework for defining synthetic indices for analyzing the structure and dynamic of scientific knowledge production \cite[e.g., ][]{bettencourt_evolution_2011, bettencourt_scientific_2009, boschma_proximity_2005, calero-medina_combining_2008, fagerberg_innovation_2009,frenken_spatial_2009, funk_dynamic_2017, grauwin_mapping_2011, grauwin_mapping_2011, rafols_diversity_2010, stirling_general_2007, uzzi_atypical_2013, petersen_grand_2021}. Building up on these efforts, here we focus our analysis around the dynamics of two complementary characteristics -- the \emph{diversity} and the \emph{relatedness} of the entities comprising the aggregate knowledge trajectory. This approach is similar to previous research using diversity measures to characterize knowledge trajectories and relatedness \cite{boschma_scientific_2014, heimeriks_how_2016, heimeriks_emerging_2012, rafols_diversity_2010,stirling_general_2007}.

To distinguish our approach to measuring diversity, we first define diversity using the typology proposed by Harrison \& Klein \cite{harrison_whats_2007}, which differentiates between three alternative perspectives: variety, separation, and disparity. Unlike previous studies, here we seek to evaluate the emergence of diversity in problem-solving approaches by measuring disparity, as opposed to variety (also referred to as richness) or separation. To be specific, while variety refers to counting the total number of varieties of entities (or richness of a system), and separation measures the characteristic differences between expressed values (differentiation), here we choose a measure of disparity because it directly measures the dominance of one or few varieties over the remaining varieties (i.e., heterogeneity in concentrations of varieties). We posit that saturation to a problem-specific diversity level is a robust indicator of whether or not a knowledge trajectory is confounded by conceptual and/or solution uncertainty. 

To develop this assessment framework, we systematically analyze disparity levels for three research areas -- Deforestation, Invasive Species, and Wildlife trade -- motivated by the following postulations:

\begin{itemize}
\label{prop:1}
\item [P1.] \textit{Invasive Species.} Given the clear task and conceptual definition of what is an invasive species, we anticipate stable knowledge trajectory dynamics for both cognitive and social dimensions (i.e., closer to a tame problem). 
\item [P2.] \textit{Deforestation.} While the definitions regarding deforestation are clear, this problem suffers from a lack of effective solutions, in part owing to multiple global stakeholders, despite being a problem that is highly localized. Hence, this problem suffers primarily from solution uncertainty (i.e., Type II wicked problem). As such, we expect such conditions to generate unstable or turbulent collaboration patterns.

\item [P3.] \textit{Wildlife Trade.} In contradistinction to P1 and P2, we suspect that instability in both cognitive and social dynamics results from relatively high conceptual and solution uncertainty (i.e., typical of a Type III wicked problem). 
\end{itemize}

\begin{figure*}
\centering{\includegraphics[width=1\textwidth]{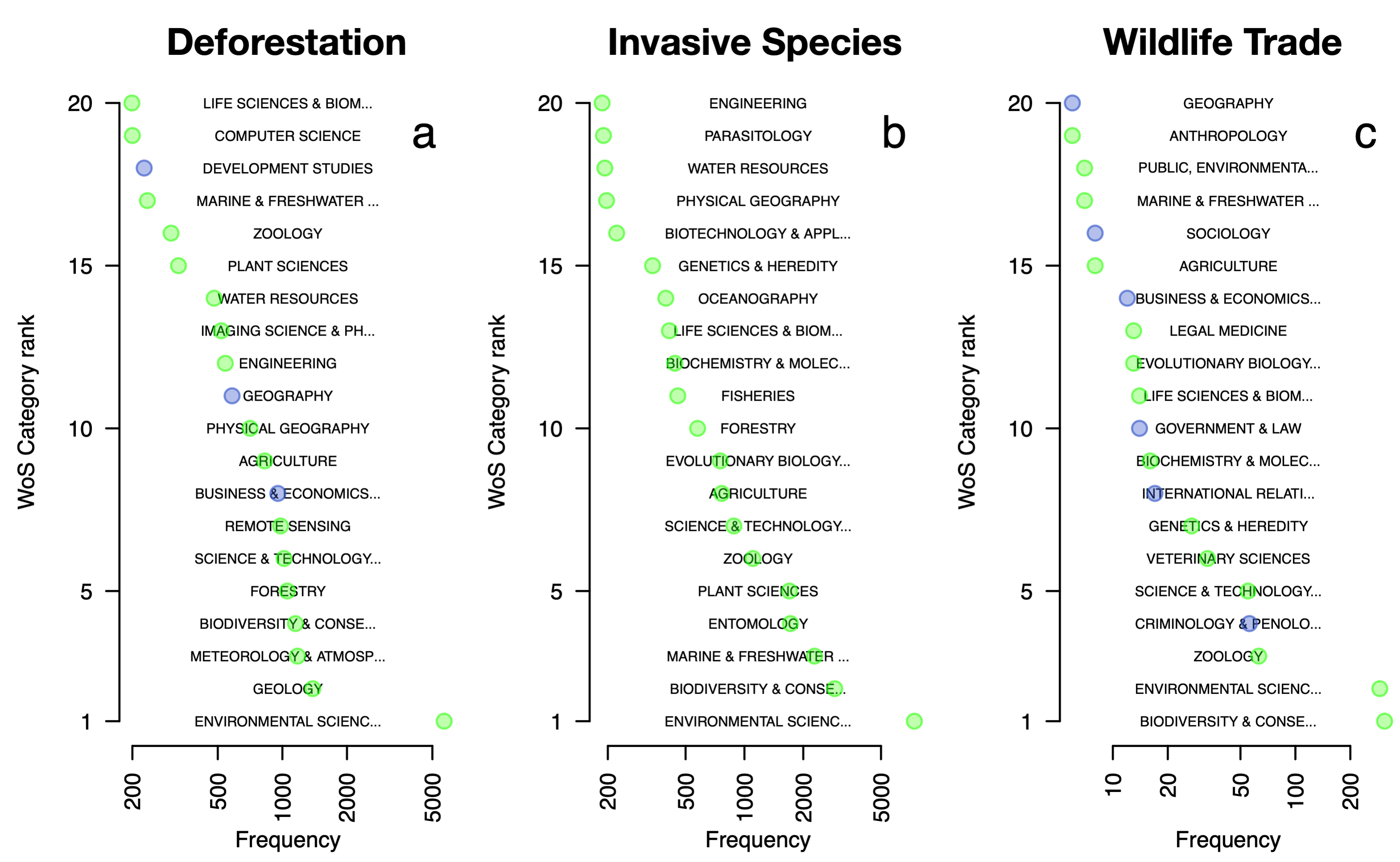}}
\caption{ \label{fig1}    {\bf Disciplinary composition of three environmental problems.}Disciplinary composition of the three environmental problems domains: (a) Deforestation; (b) Invasive Species; (c) Wildlife Trade. Shown are the top 20 most frequent WoS categories associated with each. Data point colors indicate physical sciences (green) and social sciences and humanities (blue).}
\end{figure*}

\vspace{-0.3in}
\section{Methods}
\vspace{-0.1in}
\label{sec:3}
We analyze knowledge trajectory change by assessing structural changes in diversity (disparity) among the constituent components of the research corpus in and around each problem domain \cite{bettencourt_evolution_2011, fagerberg_innovation_2009, heimeriks_how_2016}. In what follows we first detail the environmental problems addressed and then we describe our proposal for assessing structural changes in a given research domain, which is sufficiently general to be applied beyond the three case studies explored in this work.

\vspace{-0.2in}
\subsection*{Environmental problems}
\vspace{-0.1in}

Grand environmental challenges involve high degrees of uncertainty in cognitive, social, and technical dimensions. By way of example, a conservation biologist may have to make decisions or recommendations about ecosystem management before a complete theoretical, empirical or methodological foundation have been established \cite{alford_wicked_2017, head_wicked_2008,head2019forty,soule_what_1985,soule_what_1998}. Therefore, tolerating epistemic uncertainty in terms of what the best available knowledge may be an unavoidable component of environmental science \cite{defries_ecosystem_2017, stirling_keep_2010}.

\vspace{-0.2in}
\subsubsection*{Three environmental problem examples}
\vspace{-0.1in}
We focus on three environmental problems of global extent (Deforestation, Invasive species, Wildlife trade) with origins in human development \cite{van_uhm_social_2018}. Since the late 1970s several studies have suggested that the three problems are both drivers and symptoms of global change, biodiversity loss and the asymmetric relationship between the global North and South \cite{barney_global_2011, scheffers2019global,willcox_history_1974}. These problems are therefore incorporated into political actions through international conventions and accords that deal with the interconnected nature and boundary crossing aspects of the phenomena at hand; examples include the Convention on International Trade in Endangered Species of Wild Flora and Fauna (CITES) of 1973, the Convention on Biological Diversity (CBD) of 1992, and the Reducing Emissions from Deforestation and Forest Degradation (REDD) of 2008.

The first problem studied is Deforestation, which refers to the intentional reduction of forest cover in both legal and illegal contexts. Deforestation has been tied to the expansion of commercial and subsistence agriculture frontier, legal and illegal logging for paper and industrial hardwood, urbanization, desertification, and climate change \cite{barbier_trade_1994, malhi_climate_2008, willcox_history_1974}. The impacts of deforestation on vulnerable populations can be wide ranging and degrade human wellbeing \cite{barbier_trade_1994,carrasco_2017}. Figure {\bf \ref{fig1}a} shows the disciplinary composition of the scientific research on deforestation, illustrating a background context involving both the natural sciences (in the endeavor to assess land cover change and its impacts), as well as the social sciences (relating to forest/agriculture management, as well as efforts to understand the sociocultural and economic impacts of deforestation). 

The second problem studied is Invasive Species, which refers to biological invasions or the unnatural demographic growth of species. Invasive species are frequently nonnative species introduced to an ecosystem either intentionally (e.g., in an active, deliberate manner) or unintentionally (e.g., in passive, accidental manner), though some native species can also become invasive \cite{cassey_global_2004, hulme_trade_2009}. The mechanisms and consequences of biological invasions differ across species, organisms, and economic settings \cite{hulme_trade_2009}. The economic impacts of controlling or coping with existing, or preventing new, invasions are significant, frequently exceeding hundreds of billion dollars per year \cite{mooney_evolutionary_2001}.  Biological invasions are mostly human driven, though ecologically shaped and filtered which reflect the disciplinary composition of the research in this problem that is notably focused on biological sciences, in particular zoology {\bf Fig.\ref{fig1}b}. From a disciplinary perspective, biological invasion is mostly addressed through the lenses of natural sciences, with little social sciences imprint despite the known consequences of invasive species in the livelihoods and economy of the inhabitants of the recipient ecosystems \cite{shackleton_role_2019}.

Finally, the third environmental problem is Wildlife Trade, or alternatively wildlife trafficking, which refers to the legal and extralegal commercialization and use of wild fauna and flora, as well as their derived products. Both legal and illegal wildlife trade frequently suffer from fuzzy boundaries that are highly debated in academia and practice \cite{challender2021mis}. Wildlife trade spans through local and international scales encompassing complex social networks that supply the increasing demand for medicines, souvenirs, pet markets, wild meats, and cultural customs \cite{arroyave_multiplex_2020,scheffers2019global,van_uhm_social_2018}. One aspect of the problem frequently highlighted is its profound ecological and social impacts \cite{arroyave_multiplex_2020} such as biodiversity loss, corruption, and violence. In contrast to the previous two problem domains, Wildlife Trade has prominent research streams in the social sciences. More specifically, {\bf Fig.\ref{fig1}c} shows that besides biological sciences, this problem is co-dominated by human sciences such as criminology and government.

\vspace{-0.2in}
\subsection*{Data}
\vspace{-0.1in}

Multiple scientific repositories have been widely used for understanding scientific dynamics. We use Web of Science (WoS), one of the most prominent sources of indexed literature \cite{leydesdorff_global_2013}, to collect the scientific literature associated with each one of the environmental problems here studied. The information was downloaded in November 2020 using general queries designed to capture each problem (see Supplementary Note 1 -SN.1-). For each publication we extracted various metadata fields, including journal (SO), authors (AU), keywords (DE), year of publication, country of authors' affiliation (CU) and, WoS research subject category (SC, similar to WoS disciplinary category (WC), both of which are journal-specific ontologies). Furthermore, for each source we also tally two co-occurrence measures, one for co-author (C-AU) and another for co-keyword (C-DE), in which we tally the frequency of dyads, i.e. the number of publications featuring author (alternatively keywords) A \& B. 

\vspace{-0.2in}
\subsubsection*{Data refinement using co-bibliography networks}
\vspace{-0.1in}
Scientific repositories systematically compile, store, and make accessible vast quantities of information regarding scientific productivity. However, these information search and retrieval engines might be sensitive to misidentifications and synonyms. To avoid including unrelated publications within our analysis we focus on publications cognitively related with at least part of the core literature associated with each problem domain. We identified such publications by reconstructing the corresponding co-bibliography network \cite{grauwin_mapping_2011, ramirez_mobilizing_2018,velden2017comparison}.

\begin{figure*}
\centering{\includegraphics[width=1\textwidth]{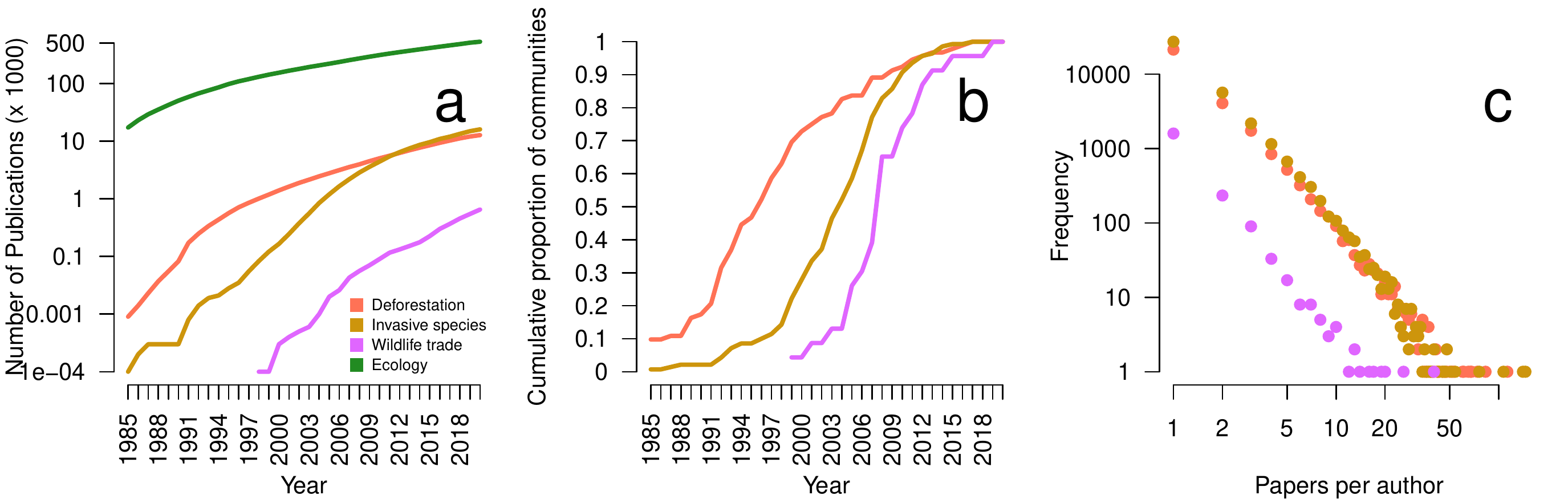}}
\caption{ \label{fig2}    {\bf General characteristics of the source sets defining the 3 case studies.}  Deforestation (orange), Invasive species (brown), wildlife trade (purple). (a) Cumulative number of sources including data for Ecology (green) for benchmarking purposes. (b) Cumulative number of knowledge communities emerging through a given year, reported as the proportion of the total in 2020 to better facilitate comparison. (c) Author productivity distribution, indicating common scaling despite underlying differences in research domain size.}
\end{figure*}

Co-bibliography networks (CBN) synthetize the association of a pool of publications through the literature they cited. CBN are composed by nodes (publications) connected by links that indicate the bibliographic similarity or bibliographic coupling between them. Publications with high similarity share a large proportion of bibliographic references, thus they are expected to address similar problems, or use similar frameworks \cite{crane_invisible_1973, palacios2018developmental}. Therefore, links in CBNs represent cognitive proximity between publications. As such, clusters of densely connected publications in a CBN represent groups of topic-specific publications. Note that we consider publications within these groups as consolidated knowledge since it underlies communities of researchers where the knowledge is discussed and diffused \cite{crane_invisible_1973, fagerberg_innovation_2009,latour2013laboratory}. While there are other extant methods for mapping publication topics, we choose CBN being that it produces maps similar to other methods \cite{velden2017comparison} and it is amenable to including recent literature that has not yet had sufficient time to be itself cited \cite{grauwin_mapping_2011}.

We assess the bibliographic coupling between pairs of publications by using the bibliographic coupling distance$^{2}$ \footnotetext[2]{ Bibliographic coupling similarity $w_{ij}$ is defined by $\frac{|R_{i} \cap R_{j}|}{\sqrt{|R_{i}||R_{j}|}}$, where $R$ represents the list of references. Hence the distance between publications $i$ and $j$ is calculated as the intersection of their references, normalized by the length of both  reference lists.}  proposed by Kesser \cite{kessler1963bibliographic} and implemented by Grauwin \& Jensen \cite{grauwin_mapping_2011}, which defines the coupling as the normalized intersection of the cited references, varying from 0 (no coupling) to 1 (identical bibliographies). Following Ramirez et al. \cite{ramirez_mobilizing_2018}, we exclude from the analysis those links that represent low cognitive proximity between papers (i.e., small coupling) by defining a threshold that maximizes the formation of highly cohesive clusters of papers (i.e., network's modularity). We use the Louvain algorithm to identify these clusters \cite{grauwin_mapping_2011}, which we term knowledge communities (KC). The threshold is defined by iteratively removing the links weighted lesser than a given value, and then measuring general properties of the resulting network such as the resulting modularity, as described in S.N.2. By removing weak links, we seek to retain the maximum of information (i.e., nodes and links) while exposing the structure defined by strong links \cite{granovetter_economic_1985,petersen2015quanti} defined here as communities forming around high cognitive proximity. Consequently, several nodes might become disconnected and form small components corresponding to tangential research. Note that studies using networks frequently rely on analyzing the giant component of the network and excluding the smaller components \cite[e.g.,][]{newman2018networks}. Here we include communities larger than an arbitrary threshold of 10 nodes, which maintains our ability to capture emerging topics \cite{ramirez_mobilizing_2018}. As such, we include nascent frameworks and ideas, but exclude inconsistent, non-related, and isolated publications.

\vspace{-0.2in}
\subsubsection*{Data characteristics}
\vspace{-0.1in}
Using threshold values of 0.167 (Deforestation), 0.177 (Invasive Species), and 0.181 (Wildlife Trade), we obtain core CBN networks comprised of: 12,674 publications for Deforestation; 15,947 for Invasive Species; and 650 for Wildlife Trade. The resulting networks are highly modular (0.88 for Deforestation, 0.95 for Invasive species, and 0.85 for Wildlife trade) indicating that the communities identified are highly cohesive and well-defined. 

\begin{SCfigure*}
\includegraphics[width=0.74\textwidth]{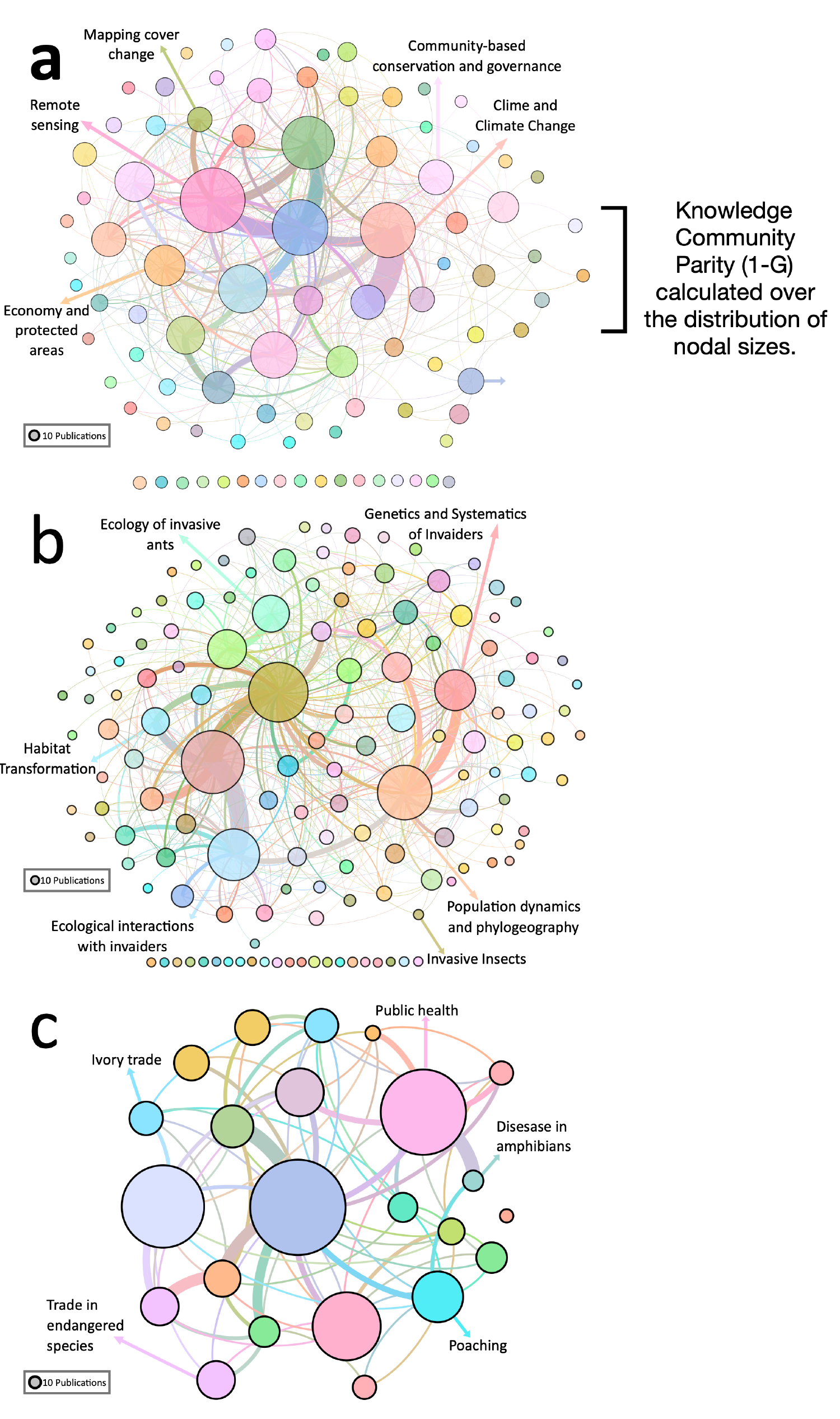}
\caption{ \label{fig3}    {\bf Cognitive composition of the three environmental problems.}  Co-bibliography networks showing knowledge communities (groups of publications represented by circles) connected by links conveying their cognitive relation (see Supplementary Note 2). We manually labeled several communities according to the topics addressed by the group. The size of the circle represents the number of publications, and the thickness of links is proportional to the number of connected publications}
\end{SCfigure*}

Importantly, we note differences in the onset of knowledge consolidation for each problem domain, indicated by the year of the first publication and the time to reach half maximum, as illustrated in {\bf Fig.\ref{fig2}a} and {\bf Fig.\ref{fig2}b}. Note that we refer to consolidated knowledge (or strongly connected CBN) rather that publications in general. Although the three problems are relatively contemporary (the earliest observation is in the 1960-70's for the three cases), consolidated knowledge for Deforestation and Invasive species emerges in the early 1980's, whereas for Wildlife Trade it emerges in the late 1990's. In Fig. S1(a,b,c) we provide additional network visualizations showing the emergence of select knowledge communities. For each case we observe a sigmoidal curve, similar to other studies of collaboration networks (Bettencourt et al., 2008), here indicating the onset of knowledge diversification since each community represents a collection of research articles that are highly coupled in terms of their knowledge inputs (see also {\bf Fig.\ref{fig3}}).

To assess the numerosity and productivity of researchers in each problem domain, we applied a simple name disambiguation method by collecting articles authored by common surname and first initial, an approach that is remarkably robust in studies of this scope \cite{milojevic_accuracy_2013}. As well documented in the literature \cite{bettencourt_population_2008, petersen_inequality_2014, sun_social_2013, wagner_network_2005}, we observe an extremely right-skewed productivity distribution {\bf Fig.\ref{fig2}c}), indicating that each problem supports just few highly productive authors, whereas the vast majority of scholars publish just few research articles. Despite the differences in the publications abundance for each problem, estimation of the skew using the single-parameter power law distribution model \(P(x=\textrm{sources per author})=x^a\), indicates similar scaling exponents (\(a=\) 2.495 for Deforestation; \(a=\) 2.494 for Invasive species; \(a=\) 2.246 for Wildlife trade). In summary, we show that the 3 case studies are not markedly different in their general characteristics, thus we argue that differences are defining features of the problem domains, as opposed to idiosyncratic differences associated with variation in sample size and scholar productivity.

\vspace{-0.2in}
\subsection*{Analytical approach}
\vspace{-0.1in}
In this section we distinguish research article metadata categories used as proxies for either cognitive or social dimensions. For the cognitive dimension we include the size of knowledge communities (KC), the frequency of keywords (DE), keywords co-occurrence or dyads (C-DE), subject categories (SC), and journals (SO). For the social dimension we include the frequency of authors (AU), coauthors dyads (C-AU) and countries (CU). Then for each variable we measure the disparity, using two measures that both correspond to greater disparity the smaller the value: Shannon Evenness index \cite{mccune_analysis_2002} and the Complementary Gini index (or simply $1-G$, where $G$ is the traditional Gini index). We measure the diversity over the publications retained in the CBN.

In more detail, Shannon evenness$^{3}$ \footnotetext[3]{The Shannon evenness $E$ is represented by $-log(m)^{-1}\sum_{i=1}^{m}log(p_{i})p_{i}$, and is a bounded version of the traditional Shannon entropy, normalized by the maximum entropy $log(m)$ associated with equally frequent ($m$) varieties; where $m$ is the total number of distinct varieties or types.} is a normalized version of the Shannon entropy which measures the average level of information contained in the variable \cite{mccune_analysis_2002}. Alternatively, the Gini index$^{4}$ \footnotetext[4]{The Gini inequality index $G$ is calculated by $(2n^{2}\overline{x})^{-1}\sum_{i=1}^{n}\sum_{j=1}^{n}|x_{i}-x_{j}|$,  evaluates the mean absolute difference between all  pairs of values ($x$),  where $n$ is the total number of data values.}  is an inequality coefficient that measures the pairwise difference between all the data values in the sample normalized by the value expected of this quantify for a uniform distribution. We use the complementary Gini (1- Gini) to simplify the comparison with Shannon. In general, these two indices measure diversity according to the disparity within the distribution of distinct varieties. Low diversity (corresponding to high disparity) is associated with a system dominated by few varieties or high homogeneity; while high diversity (low disparity) represents high heterogeneity in the system since varieties are uniformly distributed a no one dominates. Both metrics vary from 0 (representing homogeneity, low diversity, high disparity, high concentration) to 1 (heterogeneity, high diversity, low disparity, low concentration).

We evaluate the temporal changes in the diversity in two ways -- intra-annually and inter-annually. In the first case we calculate the intra-annual diversity using just the varieties that exist in each given year. Changes in intra-annual diversity indicate whether the disparity varied in a particular non-overlapping time-frame. For instance, a decrease in intra-annual authorship diversity indicates that the publications in a given year increasingly concentrated on just a few productive authors. Note that successive intra-annual diversity values might be the same, indicating for instance certain degree of concentration in a particular author, but the disparity can be produced by different varieties (e.g., author $a$ dominates in year 1 and author $b$ in year 2). 

We complement the intra-annual perspective with a second cumulative perspective on diversity change, calculated by accumulating varieties from the beginning of the data up through the specific year being analyzed. In this way, an increase in the inter-annual diversity indicates that varieties included in year $t$ were marginally represented in the past, and possibly non-existent. We argue that this inter-annual perspective accounts for the temporal change in path-dependent or relatedness of the varieties existing up through year $t$, since variations depend upon the intra-annual diversity of a given year and the previously existing varieties. In other words, the inter-annual diversity evaluates how varieties in time $t$ fit into the existing trajectory \cite{funk_dynamic_2017}.

Finally, we assess the degree to which a given disparity value could arise from random configurations of the same empirical varieties, calculated by estimating the average diversity values obtained through a random null model. In this way, the null model captures patterns representing a baseline for a particular problem, where the variables of interest have no effect \cite[see,][]{swenson2014functional}. Our null model is comprised of 5000 random ensembles, where each ensemble of varieties is obtained by shuffling the publication years (i.e., intertemporal resampling), followed by the calculation of the intra- and inter-annual diversities. We then represent the estimated null diversities as the mean diversity across 5000 realizations along with the corresponding inter-quartile range. In this way, we conserve the number of research articles analyzed in a given year, but allow for variation in their other covariates. The objective of this comparative baseline is to assess to what degree temporal patterns can be explained by phenomena in excess of the intrinsic fluctuation level associated with the entry and exit of varieties, as well as their frequency dynamics. 

Note that in cases where there is little difference between the empirical intra-annual diversity and the null model suggests that the diversity is a product of the number of varieties included, but not of their distribution. Moreover, small differences between the empirical inter-annual diversity and the null model indicates the absence of temporal sorting and sporadic bursts in the underlying data. If the temporal distribution of varieties has some order (e.g., some prolific author only published in early years and then were replace by new prolific authors) it is unlikely that the null model captures such order. Note also that in the last year of evaluation differences in the inter-annual diversity between the null model and the empiric data are not expected since the distribution of varieties in the last year is the same for both. All the calculations were made in R 3.6 \cite{team2013r} using the package igraph \cite{csardi2006igraph}.

\vspace{-0.2in}
\section{Results}
\vspace{-0.1in}
\label{sec:4}
In the previous sections we document the similarities in growth rate, author productivity distribution, and modularity of the co-bibliography networks (CBN) shown in {\bf Fig.\ref{fig2}}. In what follows, we first further describe the CBNs for each of the three environmental problems, and then we assess the disparity dynamics, used as proxies for the temporal structure of the cognitive and social dimension of the overall knowledge trajectory. Importantly, the observed dynamics are consistent regardless of the diversity measure used, as indicated by comparing calculations using Shannon Evenness (Fig.S2-S5) and Complementary Gini indices (Fig. S6-S9). Thus, we choose to focus the remainder of our analysis on the results obtained using the complementary Gini index.

\begin{figure*}
\centering{\includegraphics[width=1\textwidth]{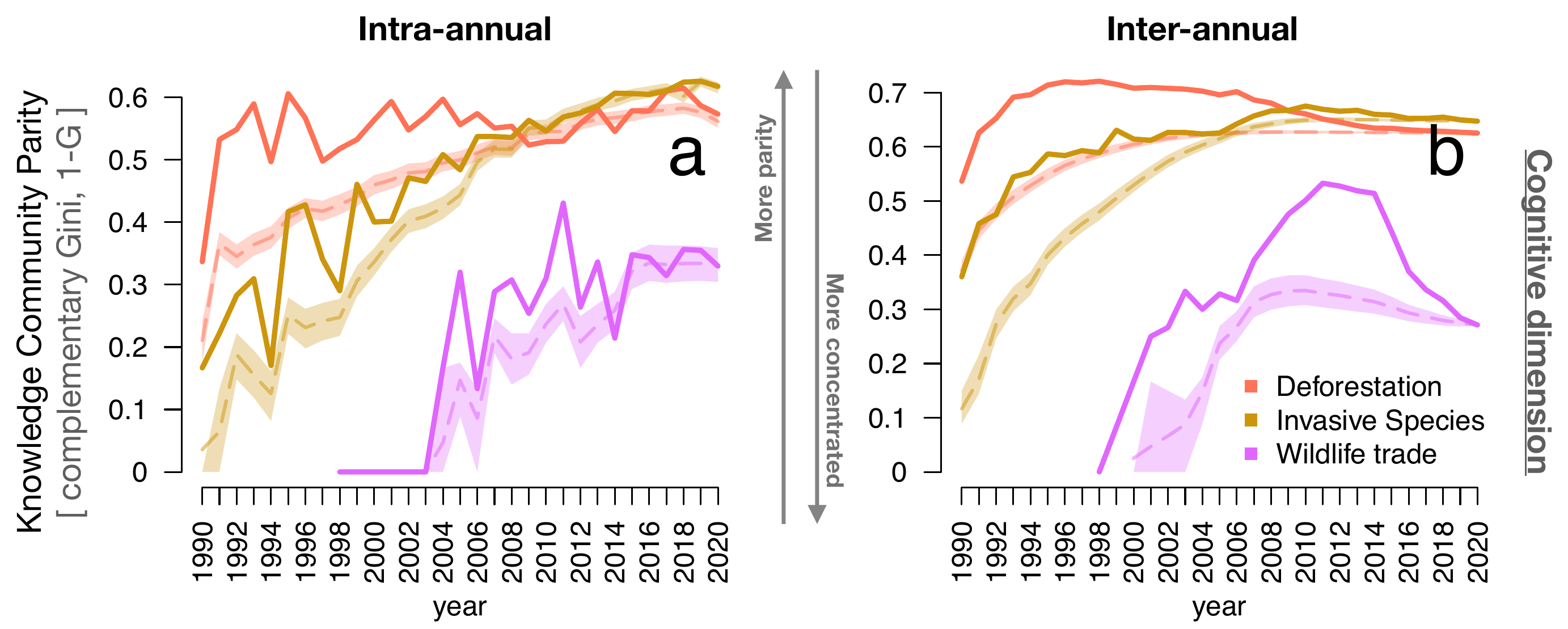}}
\caption{ \label{fig4}    {\bf Temporal variation in parity in cognitive dimension.} Temporal parity measured as the complementary Gini index (1-Gini) in knowledge community sizes for three problem domain areas: Deforestation (orange), invasive species (brown), and wildlife trade (purple). (a) Intra-annual variation. (b) Cumulative (inter-annual) variation. Larger (smaller) parity values (reported as 1-Gini Index) correspond to lower (higher) concentration levels.  Shaded intervals denote the interquartile range for data generated by randomized null model, applied to each domain separately; dashed lines indicate the mean null model realization value}
\end{figure*}

Knowledge communities (KC) are natural elements for analyzing the structure of CBNs, given their co-bibliographic construction. In {\bf Fig.\ref{fig3}} we show a simplified representation of each CBN (for detailed networks see Fig. S1) in which nodes represent KC and links between them represent the number of articles connected (representing bibliographic coupling) between the papers included in each pair of KCs. Note that the number of KC between each problem analyzed vary. Differences in the sizes of KC (number of papers included) within each problem are evident, showing that the BCNs are composed of a few very large communities and a (relatively) large collection of small communities, some of which are disconnected from the network's fully connected (giant) component. 

As mentioned, KC are clusters of publications cognitively proximal research publications. Such clusters  represent   conceptual and methodological frameworks,  developed by scholarly communities incrementally over time  for the purpose of forming a coherent scientific discourse \cite{calero-medina_combining_2008, grauwin_mapping_2011, rafols_diversity_2010, ramirez_mobilizing_2018}. By manual assessment of titles and abstracts we can identify the topics covered in each KC. For example, we find that Deforestation KC encompass a variety of topics such as the relation between land cover and water quality, fragmentation and habitat use, human impacts on habitat integrity, the role of forest in economic growth and equity, and the relationship between production of sustainable energies and deforestation, among other topics (see {\bf Fig.\ref{fig1}a}, Fig. S1a). KC topics for Invasive Species include the genetic structure of invasive populations, comparative biology between invasive and non-invasive species, management of invasions, dispersion and spatial structure of invasion, and invasions in human-dominated ecosystems ({\bf Fig.\ref{fig1}b}, Fig S1b). Finally, Wildlife Trade is characterized by KC topics associated with criminology, invasive species derived from wildlife trade, epidemiology and public health, the relationship between wildlife trade and social media, law enforcement and policy, among other topics ({\bf Fig.\ref{fig1}c}, Fig. S1c). Note that the collection of topics included in each problem reflect the multiple views that researchers develop. Although many views can be complementary, it is likely each one emphasizes specific elements (concepts) of the problem, and therefore also addressing possible solution pathways.

\vspace{-0.2in}
\subsection*{Cognitive Dimension}
\vspace{-0.1in}

We analyze the cognitive dimension of each knowledge trajectory by assessing the changes in the disparity -- for both intra and inter-annual levels. First, {\bf Fig.\ref{fig4}a} shows the intra-annual variation for knowledge community sizes, which indicates a sustained increase in diversity (increasing parity) across all problem domains. However, we note for Invasive Species and Wildlife Trade that the trajectories are only slightly greater that the expected values yielded by the null model. This suggests that the intra-annual diversity for these two problems is consistent with the random expectation and the changes in the diversity are the product of the increase in the volume of publications. In contrast, the KC diversity for Deforestation prior 2008 can't be explained by the abundance of publications, suggesting the existence of some internal process associated with the temporal distribution of efforts across multiple KC that lead such high initial diversity.

On the other hand, analysis of inter-annual variation, which better accounts for inter-temporal correlations manifesting in burstiness, shows a rapid increase in the diversity with large deviations from the null model, especially during early periods ({\bf Fig.\ref{fig4}b}). However, Wildlife Trade, and also Deforestation to a lesser degree, feature prominent decreases in diversity corresponding to higher concentration levels (i.e., smaller Complementary Gini index values). Additionally, complementary Gini index values for Wildlife trade are consistently smaller than the rest of the problems, indicating in general a lower baseline diversity, which is indicative of higher concentration of knowledge within certain KC. We posit that both saturation around a stable value, as well as higher concentration levels, can be associated with higher relatedness. In such a case, as more recent publications are incorporated, they tend to disproportionality contribute to the growth of a few existing KC, as opposed to creating new KC or being homogeneously distributed across existing KC. As such, results for KC indicate that topical diversity emerges through diversification of varieties which tend to grow equitably towards a stable saturation point at which point the system of knowledge is coherently related, as it is the case of Invasive Species. 

\begin{figure*}
\centering{\includegraphics[width=1\textwidth]{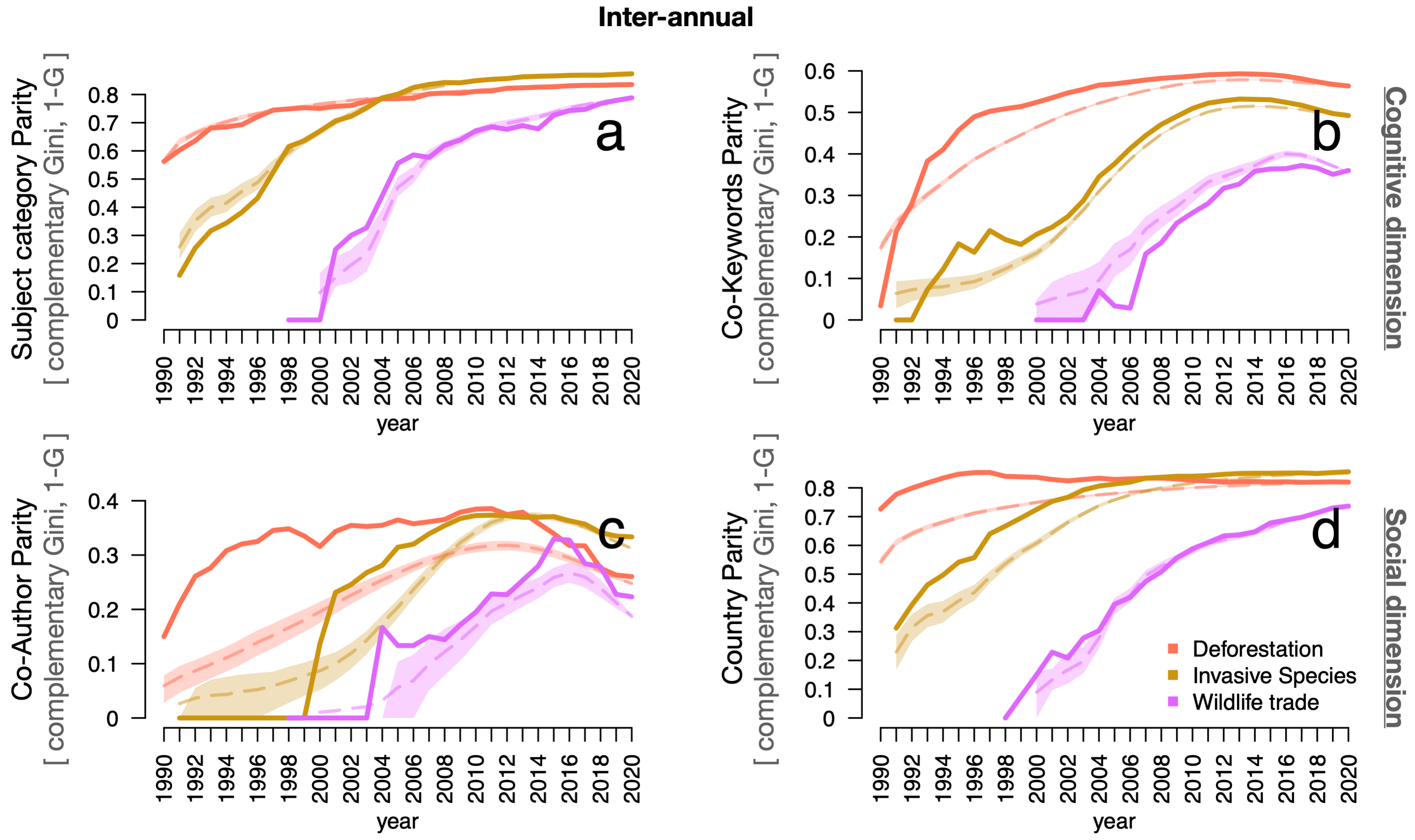}}
\caption{ \label{fig5}    {\bf Temporal variation in parity in cognitive and social dimensions.}Cumulative (inter-annual) parity calculated for three problem domain areas -- deforestation (red), invasive species (blue), wildlife trade (purple) -- and 4 different types of cognitive and social networks: (a) WoS Subject categories; (b) co-keywords dyads, (c) co-authors dyads, and (d) author affiliation country. Shaded intervals denote the interquartile range for data generated by randomized null model, applied to each domain separately; dashed lines indicate the mean null model realization value. For comparison, intra-annual analysis is provided in the SM.}
\end{figure*}

In addition to KC, we also computed the disparity time series for several other cognitive dimension variables ({\bf Fig.\ref{fig5}a-b}), including subject category (SC), journal (SO), keywords (DE), and keywords dyads (C-DE); for simplicity we present only a subset of these results, and the rest are presented in Figs. S2-S9. Results for intra- and inter-annual variation in the diversity of the mentioned variables indicate that  the variables are mostly indistinguishable from the null expectation at the intra-annual level; the inter-annual variables follow a generic increase in diversity that coincides the null expectation, except for Deforestation which features a relatively high initial diversity ({\bf Fig.\ref{fig5}a-b}); Wildlife Trade features relatively high concentration levels for inter-annual dynamics ({\bf Fig.\ref{fig5}a-b}, Figs S2-S9). As such, we corroborate that the main differences in the cognitive dimension between the problem domains is in the analysis of KC disparities ({\bf Fig.\ref{fig4}a-b}). 

\vspace{-0.2in}
\subsection*{Social dimension}
\vspace{-0.1in}
In order to analyze the social dimension of each knowledge trajectory, we present results for co-author and country disparity, as indicators of the relatedness of social communities (see Fig. S2-S9 for results derived from other social dimension variables). In particular, we focus on results corresponding to inter-annual diversity. 

Figure {\bf Fig.\ref{fig5}c} shows the results for C-AU (which are nearly identical to the results obtained for the authorship variable), which indicate that collaboration is a variable that largely differs from the expectations of the null model for Deforestation and Invasive Species, but less so for Wildlife Trade. Although we identify prolific authors ({\bf Fig.\ref{fig3}c}) in the research domain of Wildlife Trade, their impact is diminished in the case of inter-annual variation where we account for their temporal ordering. Interestingly, for Wildlife Trade and Deforestation we note a reversal towards higher concentration during the past decade, indicating an increase in the social relatedness possibly owing to a greater exploitation of existing collaborations. This suggests, at least for Deforestation, that the research activity has recently concentrated in a subset of authors and their collaborators.

Analysis of  author affiliation country data provides a proxy for diversity in organizational  (e.g., Universities, Research centers, NGOs) and institutional factors (e.g., national science funding). Results reported in {\bf Fig.\ref{fig5}d} show that empirical diversity calculated for Deforestation and Invasive Species feature early excess parity with respect to null model levels. Over time these differences reduced as parity levels stabilized around steady values, indicating high geographic parity.  Wildlife Trade features little deviation from random expectation, and parity has steadily increased over time, corresponding to a decreased concentration of geographic leadership. To further support these results, we also calculated productivity diversity between the global South and North (Figs. S5i-S9i) and also observe  inequalities in the production of knowledge generally decreasing over the long run. 

In summary, we identified important differences across the three environmental problems evaluated. Invasive Species is characterized by an increasing diversity in both cognitive and collaboration trajectories that saturates in recent times. Such a pattern describes homogenous growth across the different topics embodied, and the community of researchers as well as their supporting organizations, in addition to growth supported by preexisting structures (i.e., topics and researchers) fostering the recent stabilization of both cognitive and social dimensions. 

Similarly, Deforestation is also characterized by high diversity levels, for both cognitive and social dimensions, and featuring an approach to a stable diversity level. However, for some cognitive (KC) and collaboration (CU, C-AU) variables, we observe a slight reduction in the diversity indicative of recent increase in concentration in some varieties (e.g., topics, authors). These suggest changes in the scope of the research in this problem either by increased emphasis or paucity in some topics and researchers. We also observe reduced productivity inequalities between the global North and South. 

Finally, Wildlife Trade represents the most distinct problem of the three, showing important changes in cognitive (KC) and collaboration trajectories (C-AU) characterized by strong reduction in the diversity after periods of sustained increased. Observed parity values are typically less than those observed for other problems and are indistinguishable from the null expectation in many circumstances (DE, C-DE, AU, C-AU, CU). These results indicate that the knowledge about this problem has grown disproportionally within a few building blocks (e.g., topics, countries), thereby reducing the development potential for this wicked problem domain. Moreover, comparison of intra- and inter-annual dynamics shows heterogeneous growth, indicating that the knowledge trajectory is not  consolidating into a stable core of research topics or research leaders (neither individual nor geographic).

\vspace{-0.2in}
\section{Discussion}
\vspace{-0.1in}
\label{sec:5}
We analyzed the social and cognitive dimensions of knowledge trajectories emerging around three environmental problems -- Deforestation, Invasive Species, and Wildlife Trade. Despite the common backdrop of sustainable development and conservation, we observe differences across the different problem domains that we attribute to the role of uncertainty associated with problem and solution identification. First, we note different time periods when these problems first emerged ({\bf Fig.\ref{fig2}b}) along with different subsequent total knowledge  production as indicated by publication volumes in each problem domain ({\bf Fig.\ref{fig2}a}). Together, these observations illustrate how problem prioritization \cite{ciarli_relation_2017} reinforces the role of path-dependency in the evolution of knowledge production, and consequently also affects the  time required for building a common understanding and agenda.

Second, we observe a broad spectrum of topical approaches ({\bf Fig.\ref{fig3}}), which may indicate contested spaces where assumptions and knowledge are debated \cite{bettencourt_evolution_2011,boschma_scientific_2014, funk_dynamic_2017}. In particular, Deforestation and Wildlife Trade exhibit a prominent period of decreasing parity across knowledge communities ({\bf Fig.\ref{fig4}b}). Third, we do not observe any indication that Wildlife Trade will achieve stability in the social dimension based upon the prominent decrease in parity observed in {\bf Fig.\ref{fig5}c}. Yet it remains to be seen if stability in the cognitive dimension of Wildlife Trade will spread into the social dimension, which is a potential avenue for change \cite{bammer_expertise_2020, calero-medina_combining_2008}. However, stability in the social dimension may exacerbate solution uncertainty by reinforcing echo chambers in which a few highly productive authors (or collectives) dominate the discourse. Such a situation could limit the development of alternative leading roles, hampering the cross-fertilization between researchers and organizations, and reducing progress towards second-order `deep learning' \cite{ramirez_mobilizing_2018, ramirez2020fostering,schot_three_2018}.

We posit that variability in knowledge trajectory dynamics indicate different wickedness characteristics \cite{heifetz_leadership_1994}. When comparing our results with those previously reported for more well-stablished domains \cite{bettencourt_scientific_2009, bettencourt_population_2008, bettencourt_evolution_2011,bonaccorsi2008search, boschma_scientific_2014, heimeriks_emerging_2012} we identify some marked differences in the three domains evaluated here, which we associate with the characteristic ill-definition of each problem.

First, in contrast with our initial expectation, we found Invasive Species to more closely correspond to a Type II wicked problem (i.e., conceptually definable but without clear-cut solution) since cognitive dimensions stabilize in parity, whereas social dimensions are still changing. Until such a stable community forms, it will be challenging to settle disagreements concerning candidate solutions. Second, our initial expectations for Deforestation were also short, as this problem appears to be closer to a Type III wicked problem when considering the instability of both cognitive and social dimensions. And finally, in the case of Wildlife Trade, our analysis confirms our initial expectation of a Type III wicked problems. As such, these two Type III problems suffer from disparities that negatively affect the development of an integrated research domain. 

We acknowledge that our approximation to capturing the evolution of these problem domains is incomplete. For example, our focus on disparity measures does not provide insights into knowledge relatedness through the lens of separation diversity, as reported in other work \cite[e.g.,][]{boschma_scientific_2014,heimeriks_how_2016, leydesdorff_global_2013}. In addition, while our operational framework illuminates the structure of research producing fundamental changes in each problem, it does not provide any additional indication as to how the particular pathways connecting cognitive and leadership micro-changes translate into macro-level knowledge trajectories. A better understanding of the causal channels through which these dynamics operate will be critical to steering wicked problem domains away from unconsolidated, unactionable and eventually neglected research traps.

To address these extant challenges, we developed a generalizable framework that compares the intra-annual to the inter-annual parity dynamics, as a way to illustrate the nuances associated with the growth and saturation of diversity. In particular, our analysis of inter-annual parity indicates that growth and stability are not mutually exclusive. Indeed, cognitive trajectories often follow a process of diversification followed by consolidation and increased relatedness \cite{heimeriks_how_2016, rafols_diversity_2010}, capturing the process by which multiple voices and meld and trigger a shared vision for moving forward \cite{funtowicz_uncertainty_1994, bammer_expertise_2020, calero-medina_combining_2008, heimeriks_emerging_2012, granovetter_economic_1985, reinders_role_2011,stirling_general_2007}. Contrariwise, locked-in or highly concentrated trajectories, as exhibited in the case of Wildlife Trade, can inhibit integrative, holistic, and post-normal approaches and instead may promote the emergence of conceptual  echo chambers in which disproportionally few topics are mainly discussed by a reduced subset of voices \cite{defries_ecosystem_2017, dolfsma_lock-and_2009, funtowicz_risk_1992, funtowicz_uncertainty_1994, soule_what_1985}.

As such, environmental wicked problems appear to necessitate integrated diversification \cite{bammer_expertise_2020, calero-medina_combining_2008, linkov_scientific_2014} in which multiple voices and approaches can be included while consolidation of existing research agendas and communities of expertise takes place \cite{ramirez2020fostering}. Balancing the tension associated with this paradox of cross-disciplinary integration will help distribute efforts and capabilities toward specific solutions that iterate towards addressing the underling complexity \cite{alford_wicked_2017, defries_ecosystem_2017, light_wicked_2020}. Failing to address the tension may give rise to untenable or unactionable solutions that hinder the translation of science-based solutions into societal action, particularly at the academic-industry-government interface \cite{leydesdorff_emergence_1996}, or neglected problems as in the case of some diseases \cite{sachs_end_2005, savard_insomnia_2001}. Indeed, extremely wicked problems are likely to suffer from a broader societal disregard for pursuing further action owing to the lack of or insufficient clarity or completeness regarding problem definitions and solutions. 

\vspace{-0.1in}
\subsection*{Supplementary Material}
\vspace{-0.1in}
All supplementary materials will be available upon publication  at \href{https://datadryad.org/stash/share/YtDDbFhYauc0mV5ICLH13d1WZsKpK_t1YIQEGkzPA_U}{(Link)}. 
Supplementary materials include Supplementary Note 1: Search queries; Supp. Note 2: Co-bibliography Networks construction; Fig S1: Co-bibliography networks for the 3 problem domains; Figs S2-S5: Dynamics of knowledge trajectories measured as Shannon Evenness; and Figs S6-S9: Dynamics of knowledge trajectories measured according to the Complementary Gini index.

\vspace{-0.1in}
\subsection*{Data and code availability}
\vspace{-0.1in}
Data can be accessed through Web of Science web page and the search equations in SN.1. Code will be available in the same link that the supplementary materials and upon request to the correspondence authors.

\bibliographystyle{pnas.bst}
\bibliography{ws-acs}

\end{document}